\documentclass[longbibliography,twocolumn,pra,aps,showkeys,showpacs,superscriptaddress]{revtex4-1}
\usepackage[usenames,dvipsnames,svgnames]{xcolor}
\usepackage{amsmath}
\usepackage{amssymb}
\usepackage{graphicx}
\usepackage[unicode=true,
 bookmarks=false,
 breaklinks=false,pdfborder={0 0 1},backref=false,colorlinks=true]
 {hyperref}
\hypersetup{citecolor=blue,urlcolor=blue}
\usepackage[T1]{fontenc}

\makeatletter
\usepackage{bm}
\usepackage{braket}

\newsavebox{\@brx}
\newcommand{\llangle}[1][]{\savebox{\@brx}{\(\m@th{#1\langle}\)}%
  \mathopen{\copy\@brx\kern-0.5\wd\@brx\usebox{\@brx}}}
\newcommand{\rrangle}[1][]{\savebox{\@brx}{\(\m@th{#1\rangle}\)}%
  \mathclose{\copy\@brx\kern-0.5\wd\@brx\usebox{\@brx}}}

\makeatother

\begin{document}

\title{Creating, probing, and manipulating fractionally charged excitations of fractional Chern insulators in optical lattices}

\author{Mantas Ra\v{c}i\={u}nas}
\email{mantas.raciunas@tfai.vu.lt}
\affiliation{Institute of Theoretical Physics and Astronomy, Vilnius University,
Saul\.{e}tekio 3, LT-10257 Vilnius, Lithuania}

\author{F. Nur \"{U}nal}
\email{unal@pks.mpg.de}
\affiliation{Max-Planck-Institut f\"{u}r Physik komplexer Systeme, N\"{o}thnitzer Stra{\ss}e 38,
01187 Dresden, Germany}

\author{Egidijus Anisimovas}
\email{egidijus.anisimovas@ff.vu.lt}
\affiliation{Institute of Theoretical Physics and Astronomy, Vilnius University,
Saul\.{e}tekio 3, LT-10257 Vilnius, Lithuania}

\author{Andr\'{e} Eckardt}
\email{eckardt@pks.mpg.de}
\affiliation{Max-Planck-Institut f\"{u}r Physik komplexer Systeme, N\"{o}thnitzer Stra{\ss}e 38,
01187 Dresden, Germany}

\date{\today{}}

\begin{abstract}
We propose a set of schemes to create and probe fractionally charged excitations of 
a fractional Chern insulator state in an optical lattice. This includes the creation 
of localized quasiparticles and quasiholes using both static local defects and the
dynamical local insertion of synthetic flux quanta. Simulations of repulsively 
interacting bosons on a finite square lattice with experimentally relevant open 
boundary conditions show that already a four-particle system exhibits signatures
of charge fractionalization in the quantum-Hall-like state at the filling fraction of
$1/2$ particle per flux quantum. This result is favorable for the prospects of adiabatic
preparation of fractional Chern insulators. Our work is inspired by recent experimental
breakthroughs in atomic quantum gases: the realization of strong artificial magnetic
fields in optical lattices, the ability of single-site addressing in quantum gas
microscopes, and the preparation of low-entropy insulating states by engineering an
entropy-absorbing metallic reservoir.
\end{abstract}

\maketitle

\section{Introduction}

Topologically ordered states of matter that support anyonic excitations are a
fascinating example for the emergence of intriguing properties from the interplay
of many interacting degrees of freedom \cite{Wen2013}. Moreover, it has been shown
that such states can form a platform for robust (topologically protected) quantum
information processing \cite{NayakEtAl2008}, provided one is able to create
and manipulate the anyonic excitations in a coherent fashion. Well known examples
of topologically ordered states are fractional quantum Hall (FQH) states
\cite{Stormer1999} and the closely related fractional Chern insulators (FCIs)
in lattice systems \cite{RegnaultBernevig2011, BergholtzLiu2013, ParameswaranEtAl2013}.
However, in solid-state devices even the direct observation of individual anyonic
quasiparticles is extremely difficult (see Ref.~\onlinecite{PapicEtAl2018} for a
recent proposal), not to mention their coherent creation and manipulation.

Atomic quantum gases have been considered as an alternative environment for studying
FQH physics already for more than a decade \cite{ParedesEtAl2003, HafeziEtAl2007,
JuliaDiaz2011, CooperDalibard2013}. However, for a long time the achievable effective
magnetic fields were not strong enough to observe quantum Hall physics and also
the entropies (or temperatures)
were rather high. This situation has now changed with several
experimental breakthroughs. On the one hand, strong artificial magnetic fields
and two-dimensional spin-orbit coupling were achieved in optical lattice systems
\cite{2014Goldman, 2016Goldman, 2017Eckardt, 2009Lin, 2012Struck, 2013Struck,
2015Kennedy, 2015Aidelsburger, 2015Mancini_edge, 2015Stuhl, 2017An, 2014Jotzu,
2017Tarnowski, 2017Tai, WuEtAl2016,LiuEtAl2014} allowing for the observation of
topologically nontrivial band structures \cite{2017Tarnowski,WuEtAl2016}, a
(quantized) bulk Hall response \cite{2014Jotzu, 2015Aidelsburger}, and chiral
edge transport \cite{2015Mancini_edge, 2015Stuhl, 2017An}.
On the other hand, quantum gas microscopes were established as tools for
manipulating and imaging atoms on single lattice sites \cite{2016Ott_rev,
2016Kuhr, 2015Parsons, 2015Haller_Mic, 2015Cheuk_Mic, 2015Edge, 2016Yamamoto,
2017Mitra, 2012Cheneau, 2017Mazurenko, 2017Brown, 2017MonikaDalibard}. Using
digital micromirror devices they allow for tailoring light-shift potentials
with high spatio-temporal resolution. This technique was very recently employed
to prepare low-entropy insulating states by engineering a potential that gives
rise to an entropy-absorbing metallic shell at the boundary of a small system
\cite{2017Mazurenko}.

Inspired by these developments, in this paper we address two crucial questions
regarding the possibility to realize and observe FQH physics in a quantum gas
microscope. The first question is: Provided a FCI state has been realized, what
are experimentally feasible probes that can reveal its characteristic signatures?
Here, our approach is to design schemes for creating and manipulating the elementary
excitations of the system and to probe their fractional ``charge'' (i.e.\ particle
number). This also paves the way towards studying their fractional (anyonic)
statistics in the future. The second question is: Are characteristic features of
a FCI state still accessible in small systems of just a few particles? This
question is of eminent practical importance because, in contrast to solid-state
systems, quantum gases are well isolated rather than kept at a given temperature
by their environment. This implies that the state of the system has to be prepared
adiabatically, starting from a topologically trivial ground state and passing
through a continuous phase transition into the desired FCI state \cite{HeEtAl2017,
MotrukPollmann2017} while relying on a finite-size gap in the spectrum.

In order to address these questions, we propose three different probes that can
be implemented in quantum gas microscopes, and simulate them exactly for small
systems of four particles on about fifty sites. We consider the experimentally
relevant scenario of repulsively interacting bosons on a square lattice subjected
to a homogeneous flux and focus on regimes where one expects a Laughlin-like FCI
state at the filling of $\nu = 1/2$ particles per flux quantum. Note that it is
not our aim to study fundamental properties of FCI states as such, as they have
been investigated already in a number of previous studies (see, e.g., Refs.~\onlinecite{
SorensenEtAl2005, MoellerCooper2015, HeEtAl2017, MotrukPollmann2017, GersterEtAl2017,
DongEtAl2017,ZhaoPRB} and also \onlinecite{CooperDalibard2013,GrushinEtAl2014,
2014Grusdt,AnisimovasEtAl2015, RaciunasEtAl2016} for other types of optical lattices).
We rather address how and whether it is possible to observe signatures of FCIs
in small realistic systems. The first probe is the interaction energy (which
can be obtained from measuring site occupations) as a function of the filling
factor $\nu$. As expected for a FCI, we find a pronounced minimum in the interaction
energy around $\nu = 1/2$. 
The second probe is the
accumulation of quantized fractional charge $\nu$ near engineered local defects
as a unique fingerprint revealing the localization of individual fractionally
charged excitations. Finally, as a third probe, we study the adiabatic creation
and pumping of fractionally charged excitations via the insertion of a flux
quantum through a thin solenoid localized on single plaquettes (as they can be
implemented using Floquet engineering \cite{WangEtAl2018}). Simulations of these 
probes indicate FQH physics already in the small systems studied here.

\section{Model}

We consider interacting bosons moving in a two-dimensional square lattice subjected
to an effective magnetic field. The system is described by a Bose-Hubbard
Hamiltonian, written in terms of annihilation and density operators,
$a_\ell$ and $n_\ell = a_\ell^{\dag} a_\ell$ for bosons on site $\ell$,
\begin{equation}
\label{eq:HHfull}
  H = -t \sum_{\langle \ell'\ell \rangle} e^{i\theta_{\ell' \ell}}
    a^\dagger_{\ell'} a_{\ell}
  +\frac{U}{2}\sum_{\ell} \hat{n}_\ell (\hat{n}_\ell-1) +\sum_{\ell} V_\ell
    \hat{n}_\ell.
\end{equation}
Here, $t$ describes nearest-neighbor tunneling. The Peierls phases
$\theta_{\ell' \ell}$, which encode a uniform flux $\phi = 2\pi \alpha$ and
possibly also local fluxes on certain plaquettes, can be implemented using Floquet
engineering \cite{2017Eckardt}. Whereas a homogeneous flux has already been achieved
experimentally \cite{2015Aidelsburger, 2015Kennedy, 2017Tai}, the creation of
solenoid-type single-plaquette fluxes is proposed in 
Refs.~\onlinecite{WangEtAl2018,GrassEtAl2018}.
The Hubbard parameter $U$ describes on-site interactions. We also consider energy
offsets $V_{\ell}$ on some sites, to be used for creating and/or trapping
quasiparticle and quasihole excitations.

Motivated by a recent experiment \cite{2015Aidelsburger}, we focus on the regime
$\alpha = 1/4$. In this situation, the band structure consists of four bands. 
The lowermost (which is adiabatically connected to the lowest Landau level in 
the low-flux regime)
is protected by a gap much wider than its band width. The formation of a
topological band structure was observed by measuring the Chern number
$\mathcal{C} = 1$ with convincing precision and fueled the hope for a possible
many-body experiment. For strong interactions this system is expected to stabilize
an incompressible FCI state at certain fractional fillings $\nu$, such
as a $\nu = 1/2$-Laughlin-type state \cite{SorensenEtAl2005, MoellerCooper2015, HeEtAl2017,
MotrukPollmann2017, GersterEtAl2017, DongEtAl2017,ZhaoPRB, NielsenNJP18}.

\begin{figure}
\includegraphics[width=84mm]{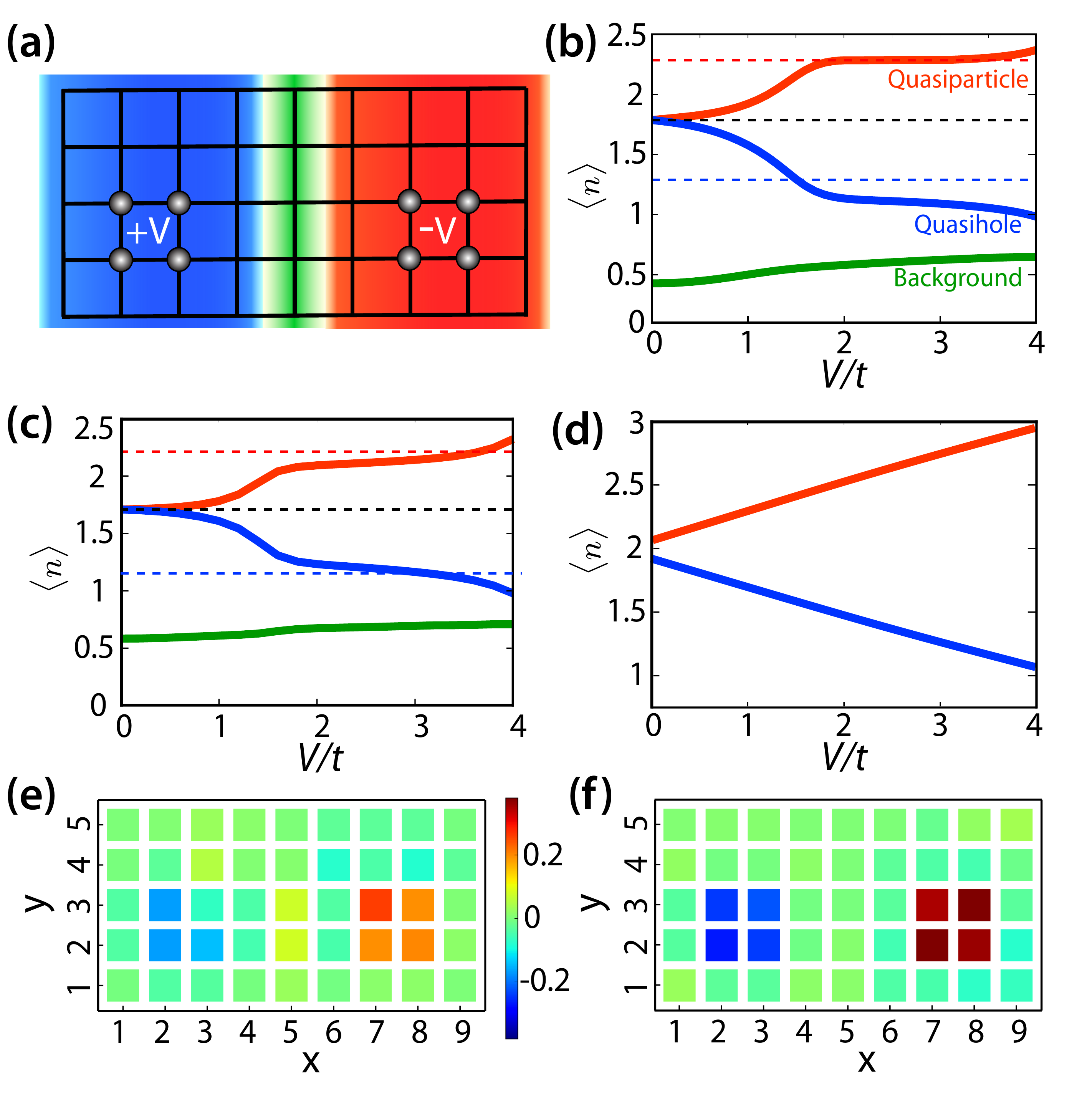}
\caption{\label{fig:charge}
Lattice geometry and charge redistribution due to additional potential offsets.
Typical geometry of $9 {\times} 5$ sites. Potential offsets $\pm V$ are each
distributed uniformly over four sites. The narrow `control' region 
in the center (green shade)
separates the left (blue) and right (red) regions where 
density change is calculated. Charge accumulation/depletion with quantized steps 
of magnitude one half is seen in lattices (b) $4 @ 9 {\times} 5$ and 
(c) $4 @ 7 {\times} 7$ operating in the topological regime; 
here $U/t = 7.5$ and $\nu = 1/2$.
(d) A trivial band insulator
leads to continuous charge flow. Density change induced by the impurity potentials:
for (e) fractional quantum Hall state at $\nu=1/2$, $V=3$, $U=7.5$ and
(f) integer quantum Hall state at $\nu=1$, $V=9$, $U=0$. }
\end{figure}

\section{Finite-size system}

In our numerical simulations, we consider a square lattice of $N_x {\times} N_y$
sites with open boundary conditions, which contains $(N_x - 1)(N_y - 1)$ plaquettes.
Such a system can be realized in current quantum gas microscope setups. To reach
the strongly interacting regime, $U\gg t$, we set $U/t = 7.5$. However, varying
the ratio $U/t$ we find little qualitative dependence on its actual value and
most of the presented results are also reproduced in the hard-core limit. The
samples are pierced by a uniform background flux, and further perturbed by either:
(i) introducing localized attractive and repulsive potential offsets, each
distributed over four lattice sites surrounding a single plaquette, or (ii)
inserting two additional localized solenoid-type fluxes of equal magnitude
and opposite signs, each penetrating a single plaquette \cite{WangEtAl2018}.
Figure~\ref{fig:charge}(a) shows a typical example with dimensions
$N_x {\times} N_y = 9 {\times} 5$. Note that the two locations of the potential
offsets or the solenoid fluxes (the affected plaquettes are marked by dots) are
well separated from each other and avoid the boundary sites. In
Fig.~\ref{fig:charge}(a) we also mark three regions in which charge redistribution
will be monitored. The broad region on the left (marked with blue shade) accommodates
the quasihole, and its counterpart on the right (red shade) accommodates
the quasiparticle. The two regions are separated by the narrow `neutral' stripe 
(green shade) that helps verify their localization.

In an infinite square lattice (or for periodic boundary conditions) the number
of sites $N_{\rm sites}$ (i.e.\ of single-particle states) matches the number of
plaquettes $N_ {\rm plaq}$. Thus, the filling factor $\nu$ is uniquely defined
as the ratio between the number of particles per plaquette $N / N_{\rm plaq}$ and
the number of flux quanta per plaquette $\alpha$,
\begin{equation} \label{eq:nu_plaq}
  \nu = \frac{N/N_{\rm plaq}}{\alpha}.
\end{equation}
Any deviation from the nominal rational value of the filling factor that corresponds
to a given FQH state results in creation of quasiparticle excitations. In a finite
system, an ambiguity arises because of extra sites lying at the system boundary.
Counting sites rather than plaquettes in Eq.~(\ref{eq:nu_plaq}) would produce
an alternative filling factor:
$\tilde{\nu} = \nu (N_x - 1)(N_y - 1)/N_x N_y$. This issue --- being of eminent
importance for the preparation of FQH states in quantum gas microscopes --- poses
a natural question: What is the optimal filling for the stabilization of a FQH droplet
in a small system?

\begin{figure}[ht]
\includegraphics[width=84mm]{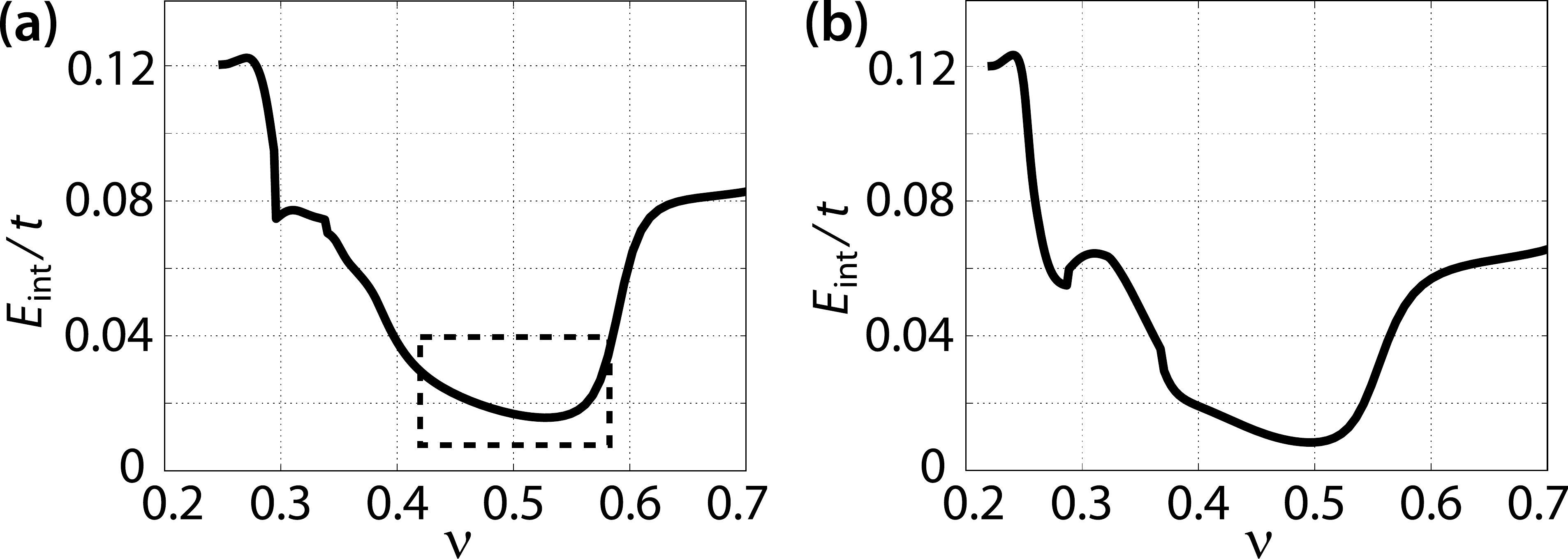}
\caption{\label{fig:flux} Scaled interaction energy, as it can be measured in a
quantum gas microscope, versus filling factor, for configurations
$4 @ 9 {\times} 5$ (a) and  $4 @ 7 {\times} 7$ (b) obtained for $U / t = 7.5$. 
The filling is tuned via plaquette flux $\phi$. Broad minima appear close to the
expected filling factor $\nu = 1/2$. }
\end{figure}

\subsection{Interaction energy}

In order to address the posed question, we consider $N$ bosons on a square lattice of
$N_x {\times} N_y$ sites (denoted $N @ N_x{\times} N_y$ hereafter). We focus on a
homogeneous system with uniform plaquette flux $\phi = 2\pi \alpha$ and in the
absence of on-site potentials, i.e.\ $V_\ell=0, \,\forall \ell$. In an experiment, 
a continuous control of the flux can be achieved, e.g., by implementing the moving
superlattice used for the Floquet engineering of the plaquette fluxes by a digital
micromirror device \cite{2017Tai}. The contribution to the ground-state energy that 
comes from the interaction of particles should attain a minimum 
at fractional filling factors that correspond to the formation of the FCI 
state. This approach was introduced for interacting fermions \cite{Kliros1991} 
on finite lattices with periodic boundary conditions, and conspicuous minima at 
odd-denominator fractional fillings were demonstrated.
We define the average interaction energy as
\begin{equation}
  E_{\rm int} = \Big\langle\Psi_{\rm gs} \Big| 
    \frac{U}{2}\sum_{\ell} \hat{n}_\ell
(\hat{n}_\ell-1) \Big| \Psi_{\rm gs}\Big\rangle,
\end{equation}
where $|\Psi_{\rm gs}\rangle$ denotes the ground state obtained from exact
diagonalization.

We begin with our principal $ 4@9{\times} 5$ configuration, which gives $\nu = 1/2$ 
for $\alpha = 1/4$ in Eq.~(\ref{eq:nu_plaq}), and vary the filling $\nu$ by tuning 
$\alpha$. Figure~\ref{fig:flux}(a) shows that the 
broad minimum of $E_{\rm int}$ has its deepest point at $\nu \approx 0.525$ 
which is just slightly off the nominal value $\nu = 1/2$.
The alternative definition $\tilde{\nu} \approx 0.71 \, \nu$ would lead to
an obvious deviation. In Fig.~\ref{fig:flux}(b) we repeat the numerical experiment
for another lattice geometry $4@7 {\times} 7$, which is of comparable area. The
minimum of the interaction energy is again centered around $\nu = 1/2$ but would
shift when plotted against $\tilde{\nu} \approx 0.73 \, \nu$.
Note that for the parameters and protocol of Fig.~\ref{fig:flux} we do not
expect a FCI state at quarter filling. In this scenario, the filling is adjusted
by tuning the plaquette flux and $\nu=1/4$ corresponds to $\phi = \pi$ that
neither breaks time-reversal symmetry nor gives rise to separate Bloch bands with
nonzero Chern numbers.

In both Fig.~\ref{fig:flux}(a) and (b), we observe rather broad minima of width
$\sim0.1$ with respect to the filling. Below (see Fig.~\ref{fig:fluxins}),
we will see that the corresponding filling factors coincide with those intervals
of $\nu$ for which we also find the quantized fractional charge transport expected
for a FCI state. This finding suggests a robustness of the FCI state with respect
to filling, which should greatly facilitate its experimental preparation. We
attribute this effect to the fact that in our small systems, a finite fraction
of the total particle number can be accommodated in gapless edge modes without
significantly changing the bulk properties of the system (which are protected by
a gap). This effect is another advantage of considering small systems
that complements their advantage for adiabatic state preparation.

We stress that besides serving as a numerical benchmark, the interaction energy
can also be used as an experimental indicator for the formation of FQH states.
The single-site resolution of quantum gas microscopes allows for extracting both
the mean and the fluctuations of the on-site occupations via measuring their
statistics in repeated experiments. This turns $E_\text{int}$ into an
experimentally accessible quantity.

\subsection{Incompressibility and charge fractionalization}

While the minimum of the interaction energy around half filling is consistent with
a FCI ground state, it does not reveal specific signatures, such as charge
fractionalization, of this state and could also indicate, e.g., a gapped density
wave at half filling. Therefore, we propose another experimentally feasible
method to probe the fractional excitations of this system. 
The method is based on the idea to introduce localized impurity 
potentials \cite{ZhaoPRB} to pin quasiparticles or quasiholes.
For that purpose we
assume that the system is prepared in the presence of two spatially-separated
local defects in the bulk: a potential dip and a potential bump. Monitoring the
integrated particle density (``charge'') in the vicinity of each defect, we expect
two signatures for a FCI state: First, the incompressibility associated with the
bulk gap should make the system stiff against weak impurities. Second, ramping up
the impurities further, we expect the dip and the bump to attract a quasiparticle
and a quasihole, respectively, corresponding to a quantized charge accumulation
of $+\nu$ and $-\nu$.

We observe that in our situation potentials situated on a single site 
(cf.\ Ref.\ \onlinecite{ZhaoPRB})
do not constitute the optimal
choice. In fact, due to the finite extent of the quasiparticles and quasiholes,
it is preferable to work with potentials $|V_\ell| = V/4$ distributed over four
neighboring sites, as depicted in Fig.~\ref{fig:charge}(a) for the $4 @ 9 {\times} 5$
lattice. We verified that results do not deteriorate when the impurity potentials
are smeared out over more than four sites as modeled by Gaussians with a width
of a lattice constant.

We calculate the ground state as a function of $V$ and plot the charge accumulated
in the three control regions [indicated by the background colors
in Fig.~1~(a)] in Fig.~1~(b).
We see that as the potential strength $V$ grows from $0$ to $2t$, the number of
particles in the left (right) end of the elongated lattice gradually grows
(decays) from the initial value of around $1.8$ to a stable value of $2.3$ ($1.3$)
(depicted with dashed lines) and remains pinned to this value in a broad region of
potential strengths up to $V \approx 4t$. We note, that the initial region of the
curves also displays insensitivity against weak potentials. Thus, we make two
observations that point to the successful creation of fractionally charged
excitations in our small system: not only the transferred charge equals $\pm 1/2$
with a good accuracy but also the initial and the final states display rigidity
typical for incompressible FCI states. The small deviation from the expected
quasihole charge of $-1/2$ can be attributed to a finite-size effect.

To further corroborate the observations, we show the results of analogous
simulations in panels (c) and (d) of Fig.~\ref{fig:charge}, respectively, for a
sample of different shape ($7 {\times} 7$) and for a topologically trivial system
(with four Bloch bands created by a superlattice potential rather than by magnetic
flux). In the former case the results reproduce the features seen for lattices
$9 {\times} 5$ while in the latter case we see a featureless uniform growth of
the transferred charge.

In Fig.~\ref{fig:charge}(e), we plot the density change induced by local potentials
of strength $V=3$ [considering the configuration of Fig.~\ref{fig:charge}(a) and (b)].
For completeness, we have also plotted the density change found for an integer
quantum Hall state on the same lattice but with $\nu=1$ and $U=0$. Here a larger
impurity $V=9$ is required to create integer charged quasiparticles/holes, since
the single-particle band gap is much larger that the many-body gap protecting the
FCI.

\subsection{Creating excitations and fractional charge pumping}

\begin{figure}
\includegraphics[width=80mm]{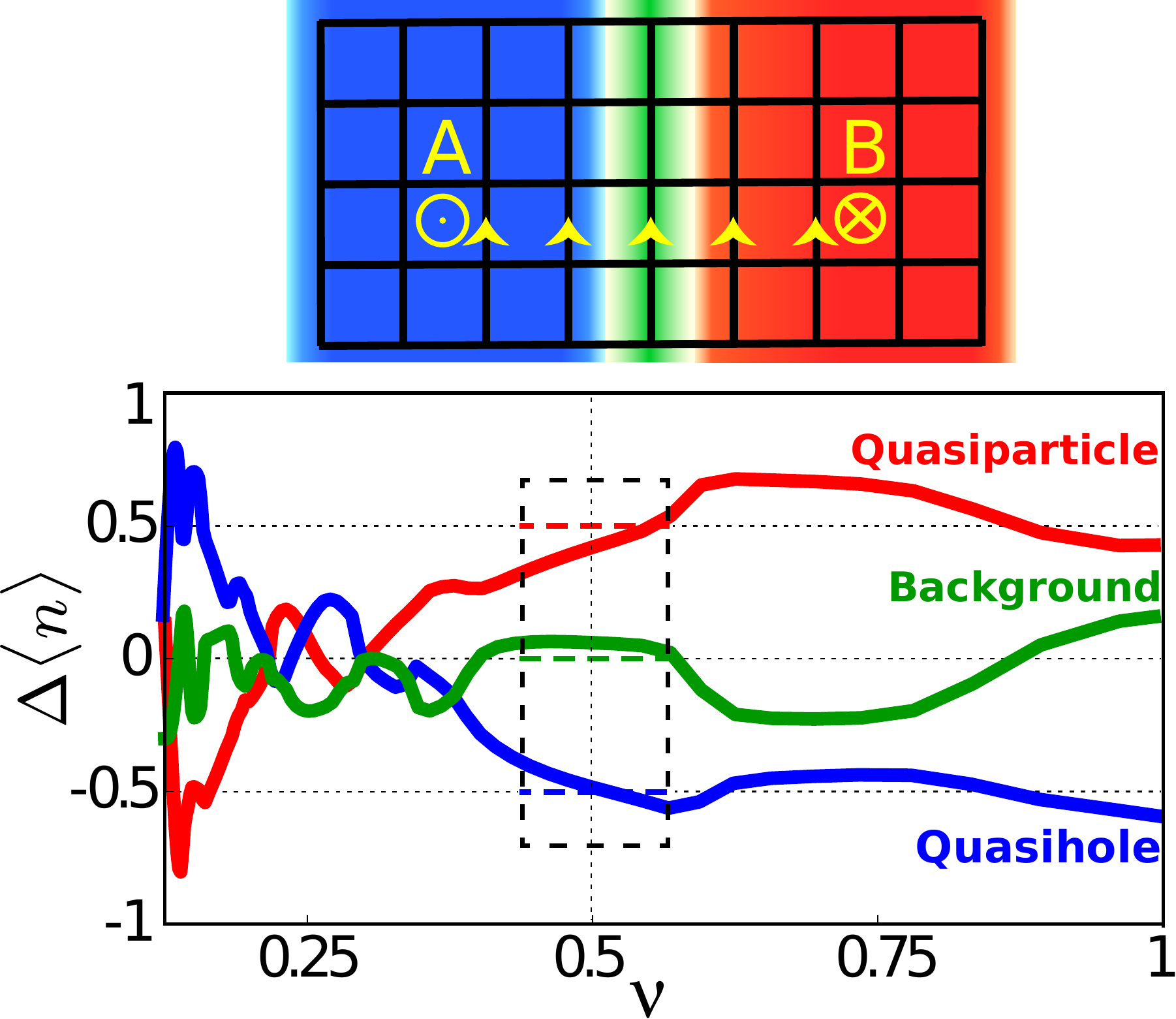}
\caption{\label{fig:fluxins}The net charge transferred in a $4 @ 9 {\times} 5$
lattice in response to the insertion of a flux quantum plotted as a function of
the filling factor. The rectangle drawn in the black dashed line marks the
vicinity of $\nu = 1/2$. The background flux is set to $\alpha=1/4$,
interaction $U/t=7.5$, and additional potential $V=\pm 0.2$ is distributed over
the A/B plaquettes.}
\end{figure}

A further hallmark of FCI states is fractionally quantized adiabatic charge 
pumping in response to the insertion of a flux quantum 
\cite{GrushinEtAl2015a,WangEtAl2018,HeEtAl2017}. 
In order to probe this effect, we consider the configuration shown in the inset of 
Fig.~\ref{fig:fluxins},
where additional Peierls phases of $\delta \phi$ are added on the bonds marked by
yellow arrows. This leads to local solenoid-type fluxes $\pm\delta\phi$ piercing
the marked plaquettes A and B on top of the homogeneous background flux $\phi$.
When $\delta\phi$ is ramped up slowly from $0$ to $2\pi$, we expect that a quasihole
of charge $-\nu$ is created at plaquette A and a quasiparticle of charge $\nu$
at plaquette B. This corresponds to the adiabatic pumping of fractional charge
$\nu$ from A to B. In order to prevent the so-created excitations from dispersing,
we also add small static local energy offsets of the same form as the ones
considered previously [Fig.~\ref{fig:charge}(a)]. We choose a very small value
$V = \pm 0.2t$, which hardly alters the ground-state charge distribution
[as can be seen from Fig.~\ref{fig:charge}(b)], but is still large enough to pin
the dynamically created local quasiparticles and quasiholes \cite{ZhaoPRB}.

For this protocol, we simulate the time evolution. Starting from the ground
state of a small system ($4 @ 9 {\times} 5$), $\delta\phi$ is switched on
linearly from $0$ to $2\pi$ moderately slowly, with the ramp time 
$\tau = 5 \hbar / t$.
We note that the ramp time $\tau$ should not be too large as the pumped 
charge would have time to disperse during the slow ramp, whereas the opposite
limit of too fast ramping would heat the system up.
In Fig.~\ref{fig:fluxins} we depict the so-induced change in the particle numbers
in the three regions indicated with different background colors in the inset.
Here the filling factor $\nu$ is again controlled by changing the background flux
$\phi$. The dashed lines correspond to the fractional charge transfer predicted
for the FCI state at half filling. We find that the pumped charge saturates
approximately to the expected value only in the region marked by the rectangular
box, where the particle density in the (green) control region also remains fixed.
In addition to being centered around $\nu=1/2$, this region matches the broad
minimum of the interaction energy shown in Fig.~\ref{fig:flux}(a) for the same
system. This provides another indication that a system as small as the one
considered here can support a FCI-type ground state.

\section{Conclusions and outlook}

In summary, we have proposed and investigated two complementary schemes for
probing charge fractionalization in optical lattice systems as a signature of
FCI states. They rely on different strategies for creating and transporting
quasiholes and quasiparticles: either by static impurities or by the local
insertion of a flux quantum. As alternative probes are not easy to implement, the
proposed density-based protocols will be crucial for measuring unique signatures
of a FCI state at a particular filling $\nu$. For example, observing the fractionally
quantized bulk Hall response to a homogeneous force is difficult because of the external
confinement, and so far, there is also no straightforward method proposed for measuring
the exchange phase $2\pi\nu$ characteristic to the anyonic quasiparticles. Also probing
chiral edge modes via the dynamics of defects \cite{DongEtAl2017} or the (experimentally
challenging) measurement of single-particle coherence \cite{HeEtAl2017} do not allow
for a clear distinction between integer and fractional Chern insulator states.
Additionally, we have pointed out that also the interaction energy is a measurable
quantity that provides further support regarding the realization of insulating
states of matter at fractional filling. Simulating these probes for a bosonic system
on a square lattice, we found that already small systems of just four particles
feature signatures of FQH state at half filling. This result is of immediate
practical relevance, since the adiabatic preparation of such a state relies on
the finite-size gap of the system.

In future work, it will be interesting to look for signatures of charge fractionalization
for FCI states at filling factors different from $\nu = 1/2$, either Laughlin states at
lower filling or hierarchy states. Another fascinating perspective is to study whether
the techniques introduced here can be extended also to measure signatures of anyonic
statistics.

\begin{acknowledgments}
We acknowledge the support from the Deutsche Forschungsgemeinschaft (DFG) via the
Research Unit FOR 2414 and from the Research Council of Lithuania under Grant 
No. APP-4/2016. It is a pleasure to thank Gediminas Juzeli\={u}nas, Julius Ruseckas, 
and Botao Wang for insightful discussions.
\end{acknowledgments}

\bibliography{droplet}


\end{document}